\documentclass[prb,aps,twocolumn,amsmath,amssymb,floatfix,superscriptaddress]{revtex4}
\usepackage[dvips]{graphics}
\usepackage{color}
\definecolor{dred}{rgb}{0,0,0.6}
\usepackage{amsmath}

\begin{document}

\title{Transport characteristics of bulk and edge states in an
off-diagonal Aubry--André--Harper chain}

\author{Moumita Patra}

\email{moumita.patra19@gmail.com}

\affiliation{School of Physical Sciences,Indian Association for the
Cultivation of Science, 2A \& 2B Raja S. C. Mullick Road, Kolkata 700032, India}

\begin{abstract}
We investigate quantum transport in an off-diagonal
Aubry--André--Harper chain. The periodic hopping modulation
generates effective internal boundaries that strongly influence the
transmission characteristics. We show that edge, in-band bulk, and
band-edge bulk states can be clearly distinguished through their
transport signatures. In particular, bulk states near the band edges
exhibit behavior similar to edge states, with weak dependence on system
size, whereas in-band bulk states display pronounced size-dependent
oscillations. We further demonstrate that the chain--electrode coupling
strength controls the broadening of transmission resonances and drives
a crossover from tunneling-dominated to nearly ballistic transport.
In addition, dephasing introduces distinct sensitivity across different
state classes, depending on their degree of spatial localization. These
results highlight the key role of internal boundaries and quantum
coherence in governing transport in modulated one-dimensional systems.
\end{abstract}

\maketitle

\section{Introduction}

Topology has emerged as a powerful framework for understanding the
robustness of quantum states against perturbations in condensed matter
systems~\cite{Hasan2010,Qi2011,Thouless1982,Kane2005a,Bernevig2006}.
In particular, topological concepts have provided deep insight into the
emergence of boundary-localized states whose properties are governed by
global characteristics of the system rather than by local details of the
lattice structure~\cite{Kane2005b,Fu2007,Moore2010}. Such boundary states
often display remarkable stability against disorder and environmental
perturbations, and they frequently give rise to distinctive transport
signatures~\cite{Hasan2010,Qi2011,Konig2007,Patra2017}. Despite these
advances, comparatively less attention has been devoted to understanding
how electron transport depends on the interplay between topological
boundary states and device-specific factors such as system size,
electrode configuration, and finite-size effects, particularly in
low-dimensional structures where boundary effects play a dominant role in
determining conduction behavior. In this work, we study the transport
properties through a topological system described by the
Aubry--André--Harper (AAH) model.

The Aubry--André--Harper model has attracted considerable
attention due to its metal--insulator transition driven by a
quasiperiodic modulation that mimics the effects of disorder in a
one-dimensional tight-binding lattice~\cite{Aubry1980,Harper1955,
Anderson1958,Kramer1993}. Typically, the modulation is introduced
through a cosine potential that is incommensurate with the underlying
lattice periodicity, leading to a localization transition at a finite
modulation strength~\cite{Aubry1980,Thouless1983,Ostlund1983}. This
prediction has stimulated extensive theoretical and experimental studies
of localization phenomena in quasiperiodic systems over the past few
decades~\cite{Sokoloff1985,Kohmoto1983,Roati2008,Modugno2010}. Notably,
experiments using ultracold atoms in optical lattices have successfully
realized the quasiperiodic AAH model and observed signatures of
localization transitions consistent with theoretical
expectations~\cite{Billy2008,Roati2008,Schreiber2015,Luschen2018}.

An intriguing aspect of the one-dimensional AAH
model is its formal connection to the two-dimensional Hofstadter model
describing electrons on a lattice subjected to a magnetic
field~\cite{Hofstadter1976,Thouless1982,Kraus2012}. This mapping
establishes a correspondence between the spectral properties of the AAH
chain and the topological features of the quantum Hall
system~\cite{Kraus2012,Lang2012}. As a result, the AAH model can host
boundary-localized states associated with gaps in the energy spectrum,
and such states have been experimentally observed in engineered
photonic and cold-atom platforms~\cite{Verbin2013,Rechtsman2013,
Lohse2016,Nakajima2016}. This connection has also enabled the
topological characterization of quasiperiodic systems and has motivated
renewed interest in studying transport properties in finite AAH
chains~\cite{Kraus2012,Mei2012}.

One advantage of the AAH framework is that it allows the investigation of
higher-dimensional topological phenomena within an effectively
lower-dimensional system~\cite{Kraus2012,Verbin2013,Ozawa2019}. In
particular, remnants of the higher dimensionality manifest as additional
degrees of freedom associated with the phase of the quasiperiodic
modulation~\cite{Kraus2012,Lang2012}. In this context, the AAH phase acts
as an effective parameter analogous to a wave vector in an additional
synthetic dimension, providing a convenient means to explore boundary and
localization phenomena within a one-dimensional
lattice~\cite{Price2015,Ozawa2019,Zilberberg2018}.

In this paper, we investigate transport in a commensurate off-diagonal
Aubry--André--Harper chain within a two-terminal open quantum
transport geometry. The system is described using a tight-binding model,
and the transmission probability is calculated within the
non-equilibrium Green's function (NEGF) formalism~\cite{Datta1995,
Datta2005,Meir1992}. We focus on a cosine modulation with period four,
which introduces an intrinsic sublattice structure; the resulting physics
can be generalized to other commensurate modulations. We systematically
analyze transport by varying system size, electrode configuration,
modulation strength, and dephasing.

Our results show that edge, band-edge bulk, and in-band bulk states
exhibit distinct transport signatures. In-band bulk states display strong
finite-size interference effects, while edge-state transport is robust
and governed by a family structure $N = 4m + r$ ($r = 0,1,2,3$), leading
to repeating boundary-dependent features within each class. Band-edge
bulk states exhibit intermediate behavior, reflecting their hybrid
character between extended and localized regimes.

We further analyze how the quasi-periodic modulation influences transport
and localization by examining the evolution of the zero-energy state as
a function of the modulation strength. This analysis provides a direct
way to characterize the role of modulation-induced internal structure in
governing localization behavior. Finally, we investigate the effect of dephasing
on transport using the
B\"{u}ttiker probe approach. This allows us to assess the robustness of
different transport regimes in the presence of phase-breaking processes
and to identify the key factors controlling coherent transport in the
modulated system.

%

The remainder of the paper is organized as follows. In Sec.~II, we define
the model and theoretical framework. Section~III presents the numerical
results and discussion of the transport properties. In Sec.~IV, we
discuss possible experimental realizations of the proposed system.
Section~V summarizes the main conclusions of this work, and
Sec.~VI contains the acknowledgements.

\section{Theory}

To model the system, we employ the tight-binding (TB) framework
within the non-interacting electron approximation~\cite{Slater1954,Datta1995}.
The total Hamiltonian of the junction is written as
\begin{equation}
H = H_{AAH} + H_S + H_D + H_T,
\end{equation}
where $H_{AAH}$ describes the AAH-modulated chain, $H_S$ and $H_D$
represent the source and drain electrodes, respectively, and $H_T$
accounts for the coupling between the chain and the electrodes.

The Hamiltonian of the chain is given by
\begin{equation}
H_{AAH}=\sum_{i} \epsilon_i c_i^{\dagger} c_i
+ \sum_{i} t_i \left(c_{i+1}^{\dagger} c_i
+ c_i^{\dagger} c_{i+1}\right),
\end{equation}
where $\epsilon_i$ and $t_i$ denote the on-site potential and
nearest-neighbor hopping integral, respectively. The electrodes are
modeled as semi-infinite one-dimensional chains with uniform on-site
potential $\epsilon_0$ and hopping integral $t_0$. The coupling
between the chain and electrodes is described by
\begin{equation}
H_T = t_S \left(c_p^{\dagger}a_{-1}
+ a_{-1}^{\dagger} c_p \right)
+ t_D \left(c_q^{\dagger}b_1
+ b_1^{\dagger} c_q \right),
\end{equation}
where $t_S$ and $t_D$ are the chain--electrode coupling strengths, and
the electrodes are connected to sites $p$ and $q$ of the chain. The
on-site potentials are set to zero throughout the analysis.

The hopping integrals follow the Aubry--André--Harper modulation
\begin{equation}
t_i = t\left[1 + v \cos\left(2\pi b i + \phi\right)\right],
\label{AAHV}
\end{equation}
where $b$ determines the modulation periodicity, $v$ is the modulation
strength, and $\phi$ is the tunable AAH phase~\cite{Aubry1980,Harper1955}.

The transmission probability is calculated using the Green's function
formalism, where the effects of the electrodes are incorporated through
self-energies~\cite{Datta1995,Meir1992}. The retarded Green's function of the chain is
\begin{equation}
G^r = \left(E - H_{AAH} - \Sigma_S - \Sigma_D \right)^{-1}.
\end{equation}
The two-terminal transmission probability is obtained from
\begin{equation}
T(E)=\mathrm{Tr}\left[\Gamma_S G^r \Gamma_D G^a\right],
\end{equation}
where $G^a=(G^r)^{\dagger}$ and $\Gamma_{S(D)}$ denote the coupling
matrices of the source (drain).

Once the transmission probability is obtained, the current is evaluated
using the Landauer formula~\cite{Landauer1970,Buttiker1986}
\begin{equation}
I(V) = \frac{2e}{h}
\int_{E_F-\frac{eV}{2}}^{E_F+\frac{eV}{2}} T(E)\, dE,
\label{but}
\end{equation}
where $E_F$ is the equilibrium Fermi energy and $V$ is the applied bias.
All calculations are performed at zero temperature.

\section{Results and discussion}

Throughout the paper, all on-site potentials are set to zero.
The nearest-neighbor hopping (NNH) integral of the chain is chosen as
$t=1\,$eV (see Eq.~\ref{AAHV}), while that of the electrodes is
$t_0=2\,$eV. We restrict our analysis to an AAH modulation period of
four sites, corresponding to $b=1/4$.

\subsection{Energy Spectrum and Localization Characteristics}

The Aubry--André--Harper model with quasi-periodic modulation has been
extensively studied as a paradigmatic system for exploring localization and
topological phenomena in low-dimensional lattices. In the present work, we
investigate the connection between the topological characteristics of the
off-diagonal AAH model and its associated quantum transport properties.
We restrict our analysis to the case where quasi-periodicity enters through
the modulation of the nearest-neighbor hopping amplitudes.

For the commensurate modulation parameter $b = 1/4$, the effective hopping
term in the system is given by
$t\left(1 + v \cos(2\pi b i + \phi)\right)$.
For a representative phase (e.g., $\phi = 0$), the cosine function takes the
values $\{0,\,-1,\,0,\,1\}$ over four successive sites, which gives rise to
four effective hopping amplitudes associated with the corresponding subbands:
$t_1 = t$, $t_2 = t(1 - v)$, $t_3 = t$, and $t_4 = t(1 + v)$.
Therefore, for $v = 1$, one of the hopping amplitudes becomes zero,
leading to a strong dimerization of the chain and the emergence of
boundary-localized states.

\begin{figure}
{\centering \resizebox*{8cm}{6.75cm}{\includegraphics{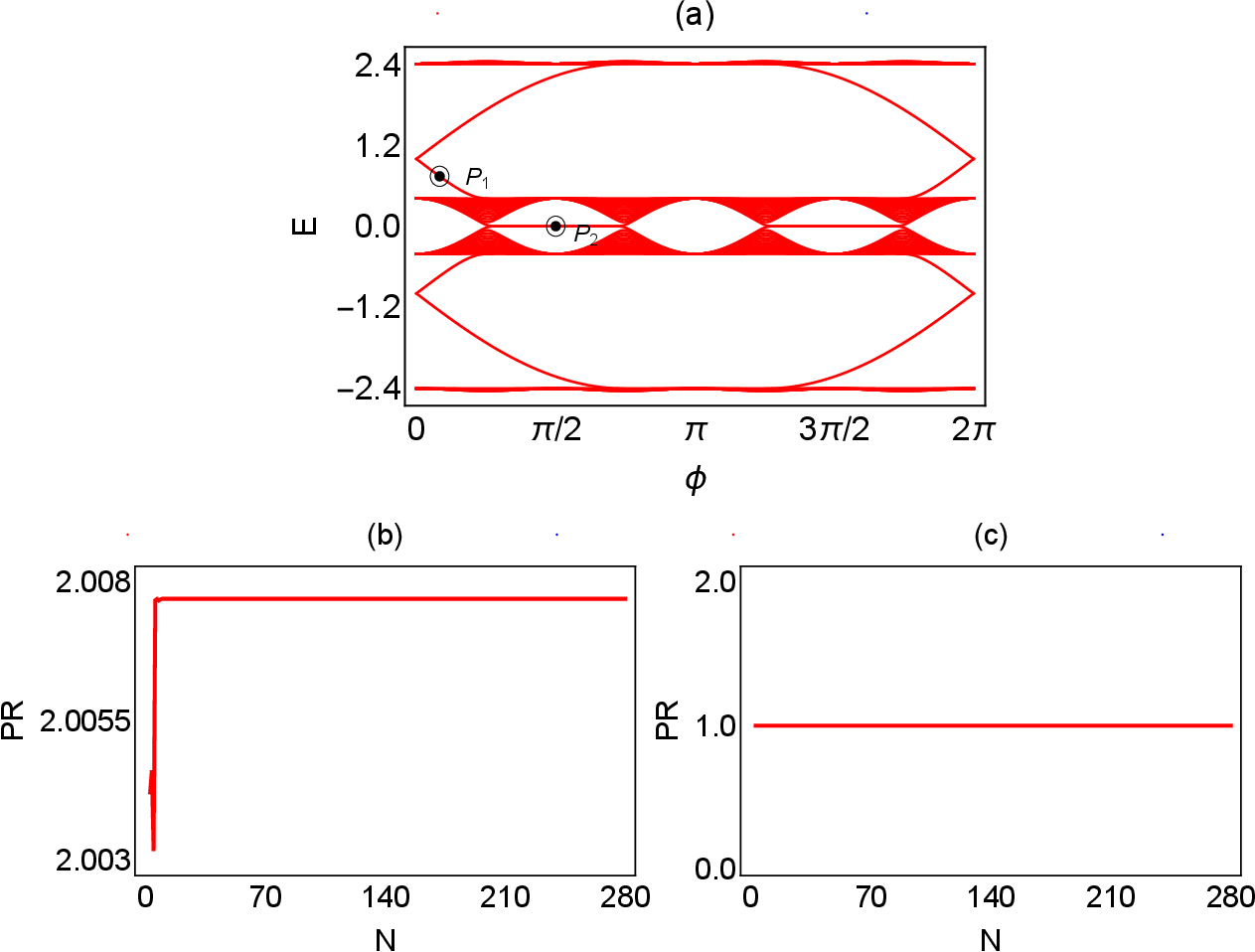}}\par}
\caption{(Color online). Energy spectrum and localization characteristics of the off-diagonal
AAH chain for $b = 1/4$ and $v = 1$.
(a) Energy spectrum as a function of the AAH phase $\phi$ for a
$60$-site chain, illustrating the formation of multiple subbands and
boundary-localized states. Two representative edge states are marked
by encircled points, labeled as $P_1$ at $E = 0.742$ and $\phi = 15^\circ$,
and $P_2$ at $E = 0$ and $\phi = 90^\circ$.
(b) Participation ratio (PR) as a function of system size $N$ for the
edge state associated with point $P_1$.
(c) Same as (b) for the flat-band edge state associated with point $P_2$.}
\label{f2}
\end{figure}

We examine the evolution of the energy spectrum as a function of the
AAH phase $\phi$ in Fig.~\ref{f2}(a), which serves as a control parameter
for tuning the band structure and gap formation. In this figure, the system
parameters are fixed at $v = 1$, $b = 1/4$, and $N = 60$. The quasi-periodic
hopping leads to a fragmentation of the energy spectrum into multiple
subbands, and the emergence of localized or boundary states depends
sensitively on the modulation strength $v$ and the boundary conditions.

The $E$--$\phi$ spectra obtained for different system sizes display a periodic
dependence on system size. In particular, the systems can be grouped into four
families according to $N = 4m + r$, where $m$ is an integer and
$r = 0, 1, 2, 3$. Within each family, the bulk-state structure of the spectrum
remains essentially unchanged, while the edge-state features are consistent
for a fixed value of $r$.

To characterize the nature of the eigenstates, we compute the participation
ratio (PR) for different energy levels. PR provides a measure of the spatial
extent of a wavefunction and distinguishes between localized and
extended states~\cite{Bell1970,Wegner1980,Kramer1993}. It is defined as
\begin{equation}
\mathrm{PR} = \frac{\left(\sum_i |e_i|^2\right)^2}{\sum_i |e_i|^4},
\end{equation}
where $e_i$ denotes the amplitude of the normalized eigenvector at site $i$,
and the value of PR represents the effective number of lattice sites over
which the state is distributed.

Here we focus on two representative edge states marked in Fig.~\ref{f2}(a) as
points $P_1$ and $P_2$. Point $P_1$ corresponds to a dispersive edge state that
connects the bulk energy bands, whereas point $P_2$ represents an edge state
\begin{figure}
{\centering \resizebox*{8cm}{10cm}{\includegraphics{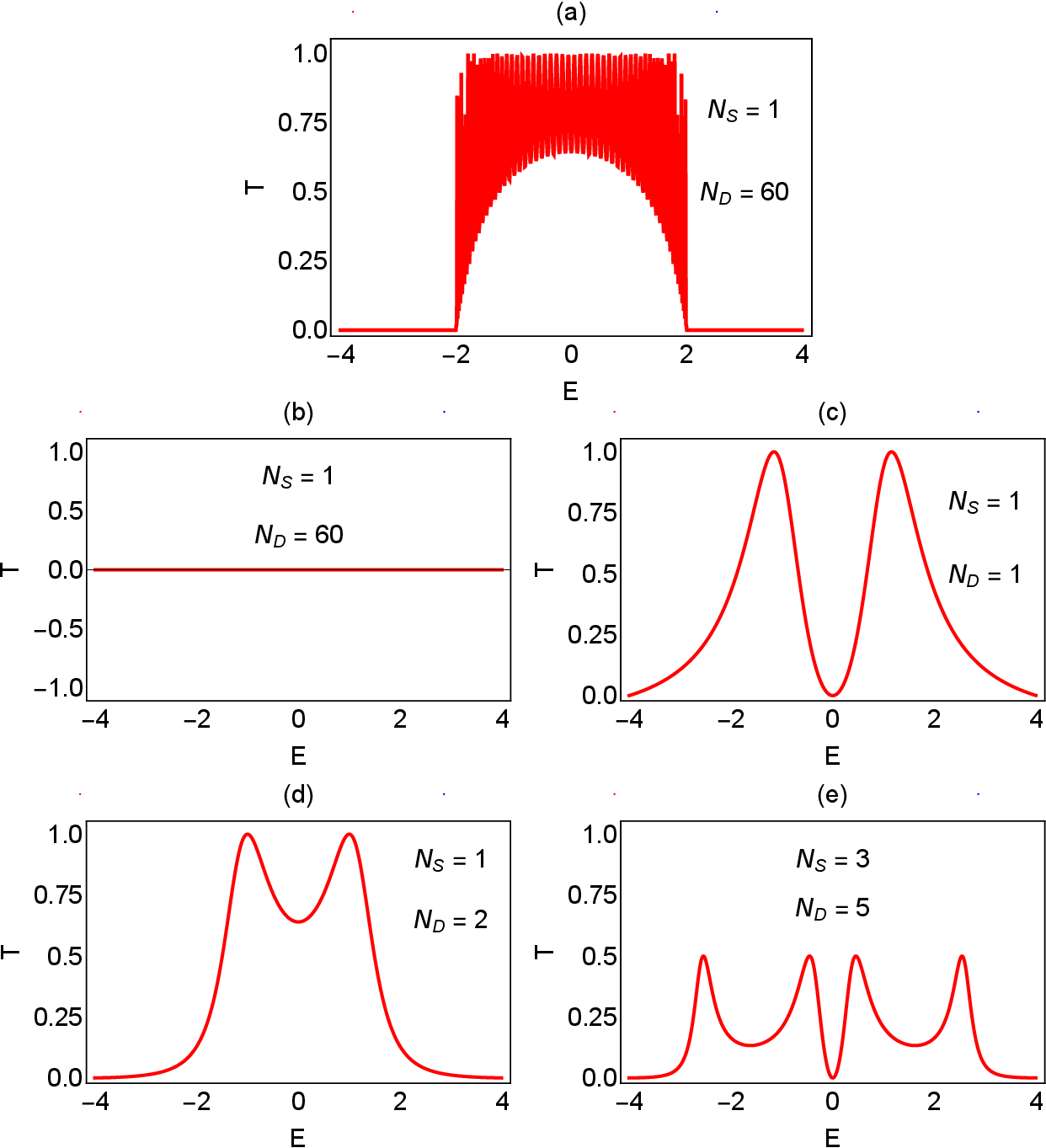}}\par}
\caption{(Color online). Transmission probability as a function of energy
for a $60$-site chain. The lead--chain coupling strength is set to
$1\,\mathrm{eV}$. (a) Ordered chain without AAH disorder, where the
source and drain electrodes are connected to sites $1$ and $60$,
respectively. (b)--(e) Transmission spectra for the chain in the presence
of off-diagonal AAH disorder with modulation strength $v = 1$,
$b = 1/4$, and phase $\phi = 0$, for different source and drain
electrode positions along the chain.}
\label{f3}
\end{figure}
associated with the flat band at $E = 0$. The corresponding system-size
dependence of the participation ratio (PR) is shown in Fig.~\ref{f2}(b) and (c).
States with relatively large PR values correspond to extended bulk states
spread over a significant portion of the chain, whereas smaller PR values
indicate stronger spatial localization.

As shown in Fig.~\ref{f2}(b), the participation ratio of the edge state
associated with point $P_1$ remains nearly constant with increasing system
size, taking values close to $\mathrm{PR} \approx 2$. This behavior indicates
that the corresponding eigenstate is localized near the boundary and extends
over only a few neighboring lattice sites, independent of the system size.

In contrast, the flat-band edge state associated with point $P_2$, shown in
Fig.~\ref{f2}(c), exhibits a participation ratio that also remains nearly
constant with system size and stays close to $\mathrm{PR} \approx 1$. This
result demonstrates that the corresponding eigenstate is strongly localized,
with its amplitude concentrated predominantly on a single lattice site.

\subsection{Transport properties of the off-diagonal AAH chain}

We now examine how the localization pattern arising from the
off-diagonal AAH modulation influences the transport properties of the
system. We begin with Fig.~\ref{f3}(a), where the transmission probability
is plotted as a function of energy for an ordered chain. In this case,
the source and drain electrodes are connected to the two extreme ends
of the chain, namely sites $1$ and $N$ ($N=60$). For a one-dimensional
ordered chain, the eigenvalues are non-degenerate, and each eigenvalue
gives rise to a distinct transmission peak. Consequently, the transmission
spectrum exhibits a series of resonant peaks within the energy band,
which ranges from $-2\,\mathrm{eV}$ to $+2\,\mathrm{eV}$ for the present
choice of hopping parameter $t=1\,\mathrm{eV}$.

In Fig.~\ref{f3}(b)--(e), we investigate the transmission spectra in the
presence of off-diagonal AAH modulation. Here we consider the modulation
strength $v=1\,\mathrm{eV}$ and restrict our analysis to the case $\phi=0$.
As shown earlier in Fig.~\ref{f2}(a), for $\phi=0$ the AAH chain exhibits
degeneracies in the energy spectrum, with bulk states appearing at
$\pm 2.41421\,\mathrm{eV}$ and $\pm 0.414214\,\mathrm{eV}$, each with a
degeneracy of $14$. In addition to these bulk states, edge states are
present at energies $\pm 1\,\mathrm{eV}$ with a twofold degeneracy.
Interestingly, the participation ratio of the degenerate bulk states is
$\mathrm{PR}\sim 2.7$, while the edge states are characterized by
$\mathrm{PR}\approx 2$, indicating localization over only a few lattice
sites.

When the electrodes are connected far apart, for example at sites $1$
and $60$, the transmission probability becomes nearly zero throughout
the entire energy window, as shown in Fig.~\ref{f3}(b). In fact, if there
is more than one atomic site separating the source and drain, the
transmission $T$ is strongly suppressed over the full energy range.
This behavior reflects the localized nature of the eigenstates in the
presence of strong off-diagonal modulation.

In Fig.~\ref{f3}(c), both electrodes are attached to site $1$, which
gives rise to non-zero transmission probabilities around
$\pm 1\,\mathrm{eV}$. The broadening of the transmission peaks arises
from the strong coupling between the chain and the electrodes. Similar
transmission peaks are also observed when both electrodes are attached
to sites $2$, $59$, and $60$, as these features are associated with the
edge states located at energies $\pm 1\,\mathrm{eV}$ with
$\mathrm{PR}\approx 2$.

Figure~\ref{f3}(d) shows the transmission spectrum when the source and
drain are connected to the first and second lattice sites, respectively.
In this configuration, finite transmission appears around
$\pm 1\,\mathrm{eV}$, with a non-zero minimum near $E=0$. This behavior
arises from the overlap of edge-state wave functions localized at
sites $1$ and $2$. An analogous result is obtained when the electrodes
are connected to sites $59$ and $60$ at the opposite end of the chain.

When the electrodes are connected to other neighboring sites, such as
sites $4$ and $5$ or $30$ and $31$, finite transmission appears
around the bulk energies $\pm 2.41421\,\mathrm{eV}$ and
$\pm 0.414214\,\mathrm{eV}$. However, this behavior is not observed for
site pairs such as $2$ and $3$ or $58$ and $59$, where the corresponding
lattice sites do not have significant wave-function overlap due to the
localized nature of the eigenstates.

The origin of this behavior can be understood from the presence of
internal boundaries in the chain, which arise at lattice sites where
the effective hopping amplitude vanishes. These internal boundaries occur
at positions satisfying the condition
$1 + v \cos(2\pi b i) = 0$. In the present case of $b=1/4$ and $v=1$,
this condition reduces to $\cos(\pi i/2) = -1$, which occurs periodically
at lattice positions $i = 2 + 4n$, where $n$ is an integer. As a
consequence, the chain becomes effectively partitioned into repeating
segments separated by these internal boundaries, leading to the formation
of localized states in the vicinity of the corresponding sites. The
resulting lattice structure can be schematically represented as
\begin{verbatim}
1  2 || 3 4 5  6 || 7 8 9  10 || 11...
\end{verbatim}
\noindent
where the symbol ``$\shortparallel$'' denotes the locations of internal
boundaries associated with vanishing hopping amplitudes.

It is important to note that although the participation ratio remains
nearly constant ($\mathrm{PR} \sim 2$), the transmission is strongly
suppressed across the entire energy spectrum when the electrodes are
connected to neighboring sites belonging to different segments, such as
$2$ and $3$, $6$ and $7$, or $10$ and $11$, due to the vanishing hopping
amplitude at the internal boundaries separating these regions.

As discussed earlier, the bulk states in this system are not fully
extended but are distributed over a limited number of lattice sites,
with $\mathrm{PR} \sim 2.7$. We further find that when the electrodes
are connected to next-to-next neighboring (NNN) sites belonging to the
same segment, finite transmission reappears at the corresponding bulk
energies. The resulting transmission characteristics are presented in
Fig.~\ref{f3}(e).

More generally, finite transmission at the bulk energy levels occurs
whenever the electrodes are placed within the spatial extent of the same
eigenstate. For example, non-zero transmission is observed between
sites $3$ and $5$, as both sites lie within the same region bounded by
adjacent internal boundaries. In contrast, the transmission becomes
strongly suppressed when the electrodes are separated by an internal
boundary, such as between sites $2$ and $4$, $6$ and $8$, or $10$ and
$12$, where the corresponding eigenstates have negligible amplitude
across the vanishing hopping link. The transmission peaks associated
with these internally localized states are typically smaller in magnitude
than those originating from the physical edge states, reflecting the
reduced spatial overlap between the localized states and the electrode
contact sites.

\begin{figure}
{\centering \resizebox*{8cm}{9.0cm}{\includegraphics{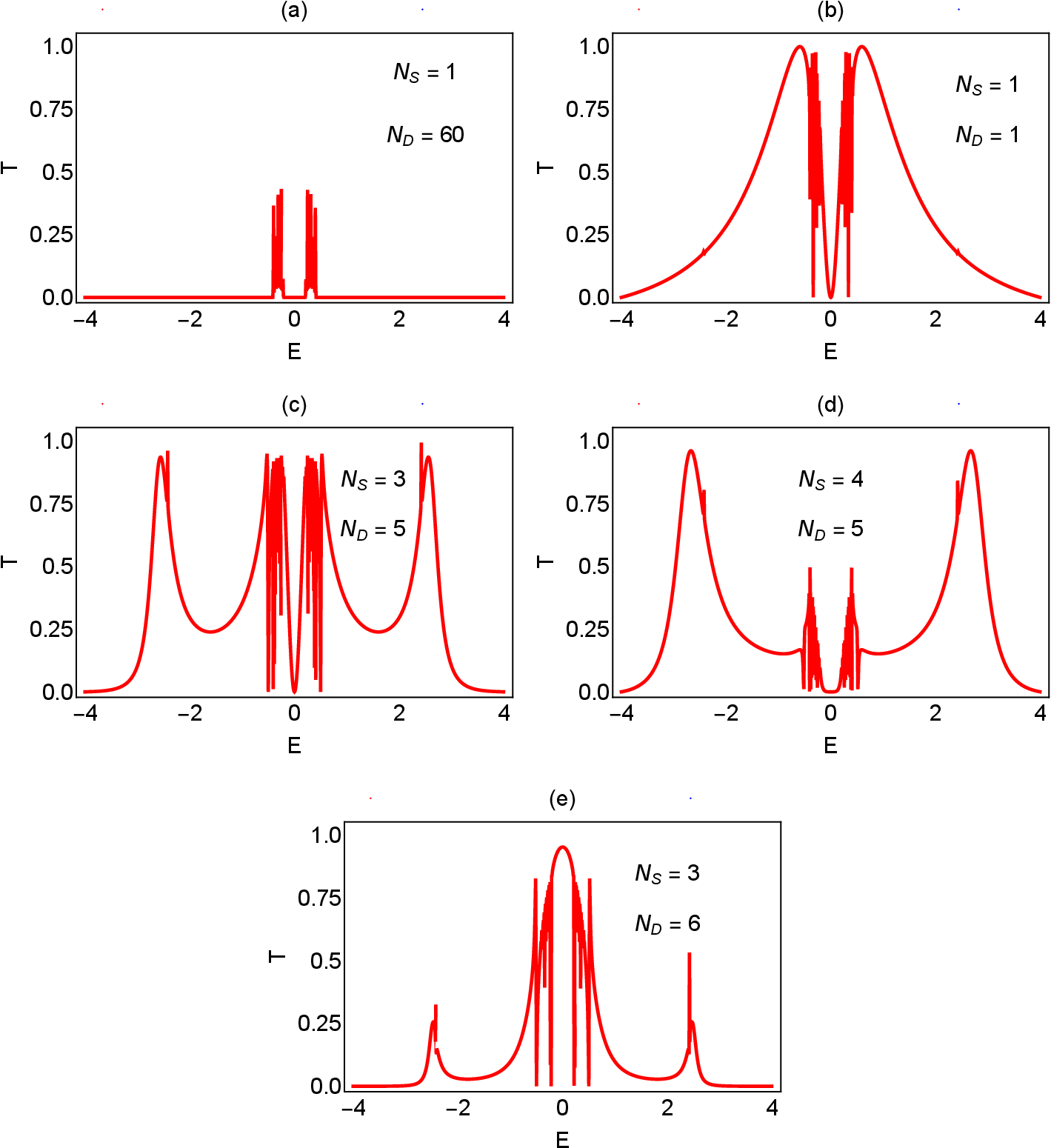}}\par}
\caption{(Color online). Transmission spectra for a chain of length $N=60$
at $\phi = \pi/6$. Bulk states are extended, with edge states at
$\pm 0.517638\,\mathrm{eV}$ and $\pm 1.50597\,\mathrm{eV}$.
Panels (a)–(e) show transmission spectra for different source and drain
positions located either in the same segment or in different segments
separated by internal boundaries.}
\label{f4}
\end{figure}

We now examine the transport spectra for non-zero values of
$\phi$, where the bulk states become extended in nature. In
Fig.~\ref{f4}, the results are presented for $\phi = 30^\circ$ and a
chain length of $N = 60$. In this case, the degeneracy of the bulk
states is lifted, and four edge states appear at energies
$\pm 0.517638\,\mathrm{eV}$ and $\pm 1.50597\,\mathrm{eV}$. The bulk
states are distributed over nearly the entire chain, resulting in finite
transmission probabilities even when the electrodes are connected to
atomic sites separated by relatively large distances.

In the present spectrum, the bulk states around
$\sim \pm 2.4\,\mathrm{eV}$ are located near the band edges, whereas
the states at $E = 0$, more precisely around $\pm 0.3\,\mathrm{eV}$
in the present spectrum, represent states within the band interior.
These two groups of states exhibit distinct transport characteristics
depending on the electrode configuration and their relative position
with respect to the internal boundaries.

For example, in Fig.~\ref{f4}(a), the electrodes are connected to sites
$1$ and $60$. In this configuration, finite transmission is observed
primarily around the bulk energies near
$\pm 0.3\,\mathrm{eV}$, corresponding to states within the band
interior. However, the bulk states near
$\sim \pm 2.4\,\mathrm{eV}$, which lie close to the band edges, do not
always contribute significantly to the transmission spectrum when the
two electrode sites do not lie within the same segment defined by
adjacent internal boundaries, even though these states possess large
participation ratios comparable to those of the states near
$\pm 0.3\,\mathrm{eV}$. This observation indicates that, in addition
to the degree of spatial extension, the relative positioning of the
electrodes with respect to the internal boundaries plays an important
role in determining the transmission response.

In Fig.~\ref{f4}(b), we present the transmission spectrum when both the
source and drain are connected to the first site of the chain. In this
case, smooth transmission peaks appear around the edge-state energies,
along with rapidly oscillating peaks associated with the $14 + 14$
bulk states within the band interior near
$\sim \pm 0.3\,\mathrm{eV}$. Although the participation ratio of the
edge states is relatively small ($\mathrm{PR} \sim 2$), these states are
localized near the left boundary of the chain and therefore contribute
significantly to the transmission only when the electrodes are placed in
their vicinity. A similar transmission profile is obtained when both
electrodes are connected to site $2$. In contrast, when the electrodes
are connected to sites away from the boundary region, the transmission
spectrum is governed primarily by contributions from bulk states within
the band interior and, under suitable geometric conditions, from bulk
states near the band edges.

The contribution from the bulk states near the band edges
($\sim \pm 2.4\,\mathrm{eV}$) becomes particularly prominent when the
source and drain are connected within the same region bounded by
adjacent internal boundaries. For example, in Fig.~\ref{f4}(c), the
electrodes are connected to sites $3$ and $5$, resulting in pronounced
and relatively smooth transmission peaks near
$\pm 2.4\,\mathrm{eV}$, accompanied by weaker oscillatory features near
$\sim 0.3\,\mathrm{eV}$ arising from bulk states within the band
interior. We further observe that these smooth transmission peaks at the
band-edge energies become dominant when the electrodes are connected to
neighboring sites located between two internal boundaries, such as
$4$ and $5$, $8$ and $9$, or $12$ and $13$. The corresponding results
are shown in Fig.~\ref{f4}(d).

Although the number of bulk states near the band edges and within the
band interior is the same, their transmission characteristics differ
significantly. The bulk states located near the band edges are associated
with relatively slower spatial variation of the wave function amplitude,
resulting in smoother changes in the transmission spectrum as the energy
is varied. In contrast, bulk states within the band interior exhibit more
rapid spatial and phase variation, leading to stronger constructive and
destructive interference between neighboring resonances. Consequently,
the transmission spectrum near the band edges displays relatively smooth
peaks, whereas within the band interior it exhibits rapidly oscillating
features.

Since the participation ratio of the bulk states in this regime is
significantly larger (typically $\mathrm{PR} > 20$), the associated wave
functions extend over a substantial portion of the chain. As a result,
finite transmission persists even when more than one atomic site lies
between the source and drain, in contrast to the $\phi = 0$ case shown
in Fig.~\ref{f3}, where the bulk states are strongly localized. This
behavior is illustrated in Fig.~\ref{f4}(e), where the electrodes are
connected to sites $3$ and $6$, both belonging to the same segment
defined by neighboring internal boundaries. However, as the separation
between the source and drain is increased further, the smooth transmission
peaks associated with the band-edge bulk states gradually decrease in
magnitude, reflecting the reduced spatial overlap between the extended
wave functions and the electrode contact sites.

The bulk states located near the band edges and those within the
interior of the band exhibit qualitatively different transport behavior.
Although both types of states may possess large participation ratios,
the band-edge bulk states are characterized by relatively flat
dispersion (as shown in Fig.~\ref{f2}(a)) and therefore smaller group
velocities, which makes their contribution to transport more sensitive
to the specific placement of the electrodes and the local lattice
structure. In contrast, bulk states within the band interior have
comparatively steeper dispersion and larger group velocities, resulting
in more robust transmission features.

We also observe that in Fig.~\ref{f4}(a)–(d) the transmission exhibits a
pronounced minimum ($T \approx 0$) at $E=0$, whereas in
Fig.~\ref{f4}(e) a relatively large transmission value appears at the
same energy. It is important to note that no eigenstate exists exactly
\begin{figure}
{\centering \resizebox*{7cm}{10.5cm}{\includegraphics{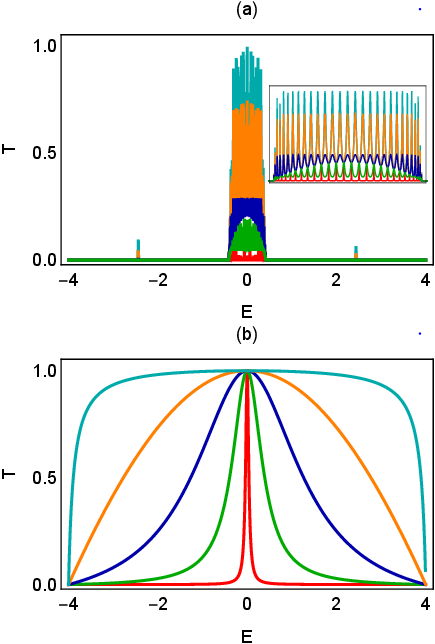}}\par}
\caption{(Color online). Transmission spectra for a chain of length
$N=60$ for different chain--electrode coupling strengths.
Red, green, blue, orange, and cyan curves correspond to
$0.2\,\mathrm{eV}$, $0.6\,\mathrm{eV}$, $1.0\,\mathrm{eV}$,
$1.4\,\mathrm{eV}$, and $1.8\,\mathrm{eV}$, respectively.
The parameters are $v = 1.0$ and $b = 1/4$.
(a) Bulk-state transport for $\phi = \pi/4$ with electrodes connected
to sites $1$ and $60$ ($N_S = 1$, $N_D = 60$). The inset shows an
enlarged view near $E=0$.
(b) Edge-state transport for $\phi = \pi/2$ with both electrodes
connected to site $1$ ($N_S = N_D = 1$).}
\label{f5}
\end{figure}
at $E=0$ for the present parameter set. The enhanced transmission in
Fig.~\ref{f4}(e) arises because the electrodes are connected to lattice
sites located in the vicinity of the same internal boundary, where the
corresponding wave functions retain significant amplitude. This
favorable spatial overlap enables efficient off-resonant tunneling
through nearby bulk states, leading to a pronounced transmission peak
even within the spectral gap.

\subsection{Effect of Chain--Electrode Coupling on Transmission Spectra}

We know that the coupling strength between the electrodes and the conductor
plays a crucial role in quantum transport, as it directly controls the
level broadening and the efficiency of electron injection into the system.
In general, stronger coupling leads to broader transmission peaks and
enhanced transmission probabilities. In this section, we investigate the
effect of the chain--electrode coupling strength on the transmission
spectra associated with both bulk and edge states.

For this purpose, we first choose a value of $\phi$ for which the bulk
states are well separated in energy, as shown in Fig.~\ref{f2}(a), and
connect the electrodes to the two ends of the chain. The corresponding
results are presented in Fig.~\ref{f5}(a). For the chosen parameters,
finite transmission occurs predominantly around $E = 0$. To better
illustrate this behavior, an enlarged view of the transmission spectrum
in this energy region is shown in the inset of Fig.~\ref{f5}(a). The
chain--electrode coupling strength is gradually increased from the
weak-coupling regime ($0.2\,\mathrm{eV}$) to the strong-coupling regime
($1.8\,\mathrm{eV}$). We find that, in the weak-coupling limit, the
transmission peaks corresponding to the bulk states are narrow and
relatively small, indicating weak hybridization between the chain and
the electrodes. As the coupling strength increases, the peaks become
broader and larger due to enhanced level broadening and stronger
coupling-induced mixing of states, and eventually approach saturation
with $T \approx 1$.

In Fig.~\ref{f5}(b), we consider $\phi = \pi/2$. From
Fig.~\ref{f2}(a), we observe that at this AAH phase a flat band appears
at $E = 0$, giving rise to two edge states localized at the two ends of
the chain. Two additional edge states are present at approximately
$\pm 2.24\,\mathrm{eV}$, localized near the left boundary of the chain,
while the bulk states remain closely spaced in energy. In this
configuration, both electrodes are connected to the first site of the
chain. As a result, the edge state at $E = 0$ primarily contributes to
the transmission spectrum because its wave function has significant
amplitude at the contact site.

In the weak-coupling limit, we observe a narrow transmission peak around
$E = 0$ with a peak height of $T \sim 1$, reflecting resonant tunneling
through the localized edge state. With increasing coupling strength,
the peak becomes progressively broader and spreads over a wider energy
range as the lifetime of the state decreases due to stronger coupling to
the electrodes. It is important to note that the energy window of the
AAH chain is $-2.41421\,\mathrm{eV} \leq E \leq 2.41421\,\mathrm{eV}$,
whereas the allowed energy range of the electrodes extends from
$-4\,\mathrm{eV}$ to $4\,\mathrm{eV}$. When the coupling strength
reaches approximately $1\,\mathrm{eV}$, the broadened transmission peak
extends toward the band edges near $E \approx \pm 4\,\mathrm{eV}$. For
even stronger coupling, the transmission profile becomes nearly flat
with $T \sim 1$ over most of the energy range and then rapidly decreases
to zero near $E = \pm 4\,\mathrm{eV}$, reflecting the finite bandwidth
of the electrodes.

These results demonstrate that the chain--electrode coupling strength
serves as an effective control parameter for tuning the transmission
characteristics of the system. It governs both the width and amplitude
of the transmission peaks associated with bulk and edge states through
coupling-induced level broadening, and drives the crossover from
tunneling-dominated transport in the weak-coupling regime to nearly
ballistic conduction in the strong-coupling limit. The overall transport
window is ultimately constrained by the finite bandwidth of the
electrodes.

\subsection{Distinct Transport Signatures of In-Band, Band-Edge, and Edge States}

In Fig.~\ref{f5}(b), we observe a single dominant transmission peak
associated with the edge state at $E=0$ across the entire conducting
energy window for a chain of length $N=60$. At first glance, it may
appear surprising that only one resonant peak is visible, since the
isolated chain possesses $60$ eigenstates. However, this peak is
robust and arises from the edge state at $E=0$, and remains
essentially independent of system size.

The eigenvalue spectra can be grouped into four
families according to $N = 4m + r$, where $m$ is an integer and
$r = 0, 1, 2, 3$. Within each family, the bulk states lie within the
same energy band, with their number increasing with system size $N$.
\begin{figure}
{\centering \resizebox*{8cm}{6cm}{\includegraphics{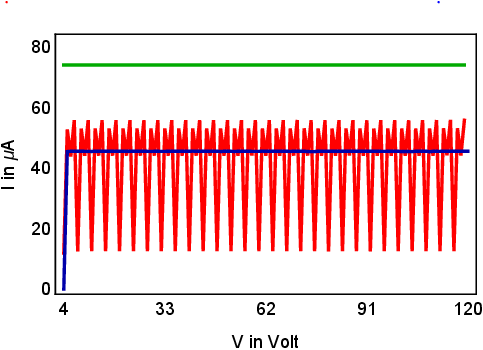}}\par}
\caption{(Color online). Output current as a function of system size $N$
for bulk and edge states at different Fermi-energy positions. The red curve
corresponds to bulk states within the band interior, obtained by setting
$E_F = 0$ with $\phi = \pi/4$ and connecting the electrodes to the two
terminal sites of the chain. The blue curve represents bulk states near
the band edge, for $E_F = 2.41875\,\mathrm{eV}$ with $\phi = \pi/4$ and
the electrodes connected to sites $4$ and $5$. The green curve corresponds
to edge states, obtained by setting the Fermi energy at the flat-band
energy $E = 0$ with $\phi = \pi/2$. All calculations are performed at
zero temperature with an applied bias voltage $V = 1\,\mathrm{V}$ and
chain--electrode coupling strength of $1\,\mathrm{eV}$.}
\label{f6}
\end{figure}
For a fixed value of $\phi$, four edge states appear for any family and
system size, and their energies remain identical within a given family.
Although slight differences may arise between different families, the
edge-state energies are essentially independent of system size within
each family. As a result, the transmission peak at $E=0$ remains
robust and effectively size-independent, as observed in Fig.~\ref{f5}(b).

To make the analysis more transparent, we examine the variation of the
output current as a function of system size $N$ for representative bulk
and edge states by placing the Fermi energy in different regions of the
energy spectrum. The current is calculated by integrating the transmission
probability $T(E)$ over energy (Eq.~\ref{but}), with the Fermi energy
positioned at the desired location corresponding to the specific bulk or
edge state under consideration. All calculations are performed at zero
temperature with a fixed bias voltage $V = 1$, and the strength of the
AAH modulation is set to $v = 1\,\mathrm{eV}$. The corresponding results
are presented in Fig.~\ref{f6}.

To investigate transport through bulk states within the band interior,
we set $\phi = \pi/4$ and position the Fermi energy within the bulk band
around $E \approx 0$. To ensure that the transmission arises primarily
from in-band bulk states, the electrodes are connected to the two extreme
ends of the chain. The resulting current (shown in red) exhibits
pronounced oscillatory behavior with increasing system size. This
behavior originates from finite-size quantum interference effects in
extended states, where the transmission probability depends sensitively
on the phase matching of propagating wave functions across the conductor.
As the system length increases, the discrete energy spectrum shifts,
leading to alternating resonant and off-resonant transport conditions,
which manifest as oscillations in the current.

In contrast, when the Fermi energy is placed near the band edge
(around $E_F \approx 2.41875\,\mathrm{eV}$), the current associated with
bulk states becomes nearly independent of system size. In this case, we
use the same AAH phase and connect the electrodes to the fourth and fifth
sites of the chain, corresponding to neighboring sites located between
two internal boundaries, where smooth transmission peaks around the
band-edge energies dominate (see Fig.~\ref{f4}(d)). Under these
conditions, the current approaches a steady value with minimal
oscillatory behavior.

A similar size-independent current response is observed when the Fermi
energy is positioned at the flat-band energy $E = 0$, corresponding to
edge states. Here we set $\phi = \pi/2$ and connect the electrodes to the
first site of the chain. These edge states are spatially localized near
the system boundaries and decay rapidly into the bulk. Once the system
size exceeds the localization length, further increases in $N$ do not
significantly modify the transport characteristics, leading to a nearly
constant current.

The contrasting behaviors observed in the three cases reflect the
underlying nature of the contributing states. Bulk states within the
band exhibit strong finite-size interference effects, whereas bulk states
near the band edge and localized edge states display robust,
size-insensitive transport characteristics. These results therefore
provide a clear transport signature distinguishing extended in-band
states from band-edge and flat-band localized states in the AAH system.

\subsection{Effects of AAH modulation parameters on transport properties}

In this subsection, we discuss the various factors that influence the
transport properties of the AAH model. For clarity, we restrict our
analysis to the off-diagonal modulation with a period of four, although
\begin{figure}
{\centering \resizebox*{8cm}{6cm}{\includegraphics{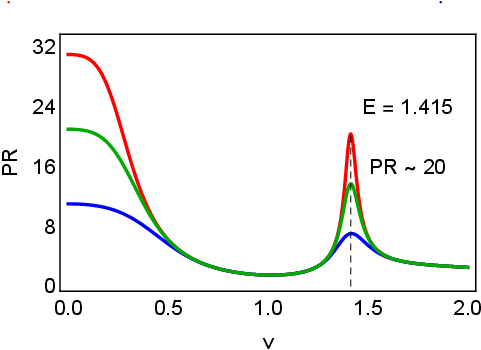}}\par}
\caption{(Color online). Participation ratio (PR) of the zero-energy state
as a function of the AAH modulation strength $v$ for $b=1/4$ and odd
system sizes $N=21$, $41$, and $61$ (red, green, and blue curves,
respectively). The nonmonotonic variation of the PR, including a local
maximum near $v \approx 1.415$, reflects a crossover from extended to
edge-localized behavior, followed by size-independent localization at
larger $v$.}
\label{f7}
\end{figure}
the underlying transport mechanism can be extended to other modulation
periods as well. For example, for $b = 1/2$, the lattice structure can be
viewed as consisting of repeating segments separated by internal
boundaries, which may be schematically represented as
\begin{verbatim}
1 || 2 3 || 4 5 || 6 7 || 8 9 || 10 11 ...
\end{verbatim}
\noindent
Within this framework, edge-state transport is obtained when both
electrodes are connected to the same lattice site at the two ends of the
chain. In particular, a terminal site can support edge-state transport
when it is not paired with its neighboring site through a vanishing
hopping amplitude that defines an internal boundary. Furthermore, if
dispersionless bulk states are present, they can exhibit edge-like
transport characteristics when the electrodes are attached either to the
same lattice site adjacent to an internal boundary or to two neighboring
sites on the same side of that boundary.

Now we investigate the effect of the AAH modulation strength on the
localization properties of the eigenstates. Varying the modulation
amplitude $v$ modifies the eigenvalue spectrum, making it difficult to
track the evolution of a particular bulk or edge state as $v$ changes.
However, for $b = 1/4$ and odd system sizes $N$, a zero-energy eigenvalue
always appears, irrespective of the value of $v$. This state therefore
provides a convenient reference point for studying the influence of the
modulation strength.

In Fig.~\ref{f7}, we plot the participation ratio (PR) of the zero-energy
state as a function of $v$ in the range $0 \leq v \leq 2$ for
$N = 21$, $41$, and $61$, shown in red, green, and blue colors,
respectively. For small values of $v$, the PR is relatively high,
indicating that the eigenstate corresponding to $E = 0$ is extended in
nature. As $v$ becomes comparable to the nearest-neighbor hopping
strength $t$ (taken as unity), the PR rapidly decreases and then
saturates over a finite range of $v$, signaling the onset of
edge-localized behavior.

With further increase in $v$, the PR exhibits a nonmonotonic variation
and shows a local maximum near $v \approx 1.415$. This feature suggests
a crossover regime in which competing hopping amplitudes temporarily
enhance the spatial extent of the zero-energy state before stronger
modulation drives it toward more pronounced localization. For larger
values of $v$, the PR again decreases and eventually approaches a small,
nearly constant value.

We also observe that when the PR is relatively large (extended regime),
its value depends on the system size $N$. In contrast, when the PR
becomes small, indicating edge-state localization, it becomes essentially
\begin{figure}
{\centering \resizebox*{8cm}{12cm}{\includegraphics{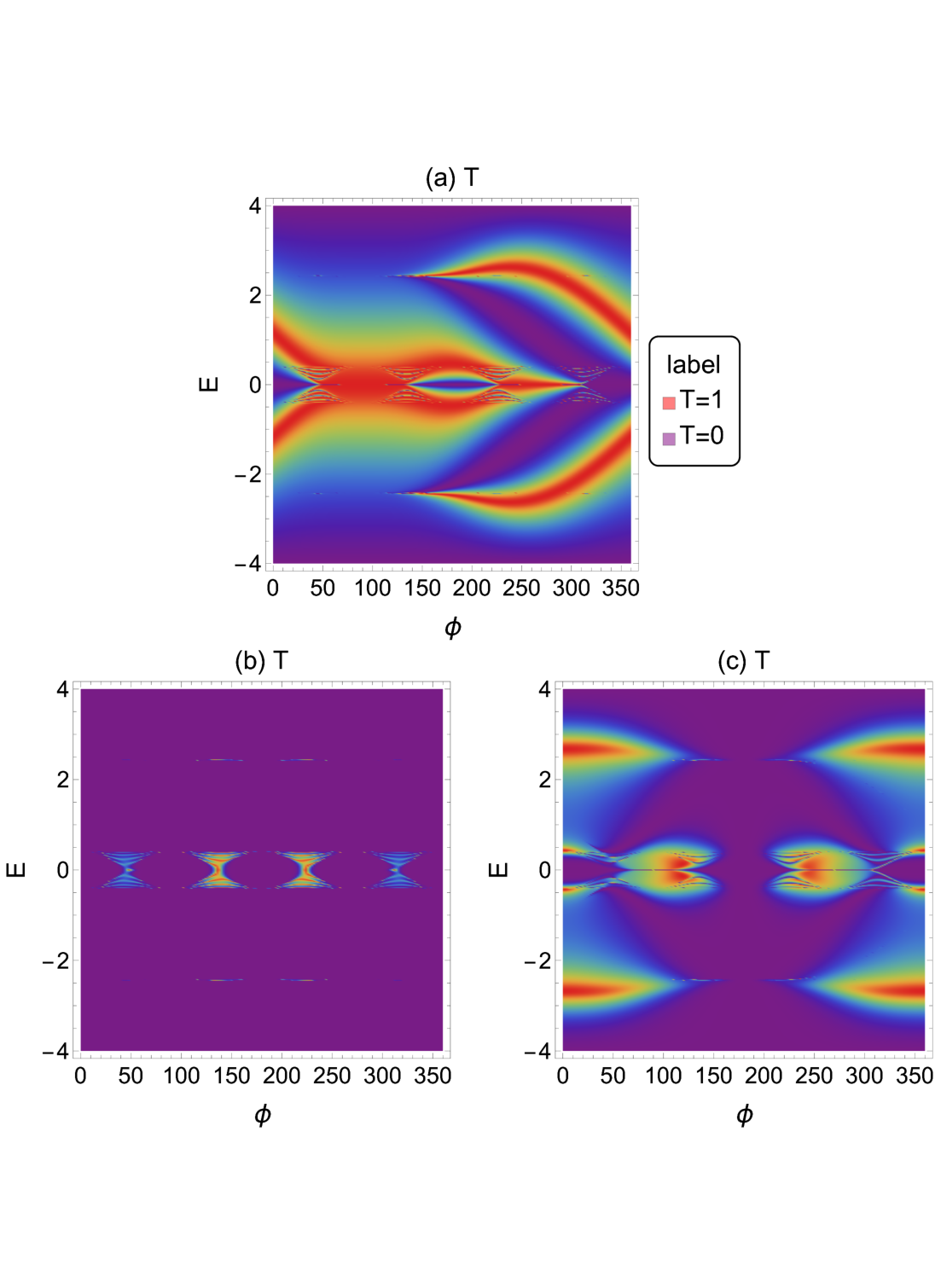}}\par}
\caption{(Color online). Density plots of the transmission probability as a function of energy
$E$ and AAH phase $\phi$ for a chain of length $N=20$.
(a) Edge-state transport with both electrodes connected to the first site.
(b) In-band bulk transport with electrodes attached to the first and
$N$th sites.
(c) Band-edge bulk transport with electrodes connected to neighboring
interior sites (fourth and fifth sites).
The qualitative features are robust against changes in system size, with
identical edge-state structures for systems within the same family
defined by $N=4m+r$ ($r=0,1,2,3$).}
\label{f8}
\end{figure}
independent of $N$, with all system sizes exhibiting nearly identical
PR values. This size-independent behavior further confirms the localized
nature of the zero-energy state in this regime.

For completeness, we present density plots of the transmission
probability as a function of energy $E$ and AAH phase $\phi$ in
Fig.~\ref{f8} for three different source and drain configurations.
The results reveal clear differences in the transport response depending
on the electrode geometry and the nature of the contributing states.
When both electrodes are connected to the first site (Fig.~\ref{f8}(a)),
pronounced and strongly phase-dependent transmission features are
observed over a wide energy range, indicating that transport is
predominantly mediated by edge-localized states. In contrast, when the
electrodes are attached to the first and $N$th sites
(Fig.~\ref{f8}(b)), transmission occurs mainly within the allowed energy
bands, consistent with transport through extended bulk states. Notably,
when the electrodes are connected to neighboring interior sites ($4$ and
$5$) (Fig.~\ref{f8}(c)), transmission features appear primarily near the
band edges with comparatively weaker intensity, indicating that
band-edge bulk states contribute to transport in a manner distinct from
conventional in-band bulk states.

All results shown here are obtained for $N=20$; however, the qualitative
features remain robust for other system sizes. While bulk-state properties
are largely insensitive to $N$, the edge-state characteristics exhibit a
systematic periodic dependence on system size, such that systems belonging
to the same family show identical edge-state behavior.

\subsection{Effect of dephasing on transport properties}

In this subsection, we examine the influence of phase-breaking processes
on the transport characteristics of edge, band-edge bulk, and in-band
bulk states in the off-diagonal AAH chain using the B\"uttiker probe
approach.

Dephasing plays a crucial role in destroying the phase coherence of
charge carriers and thereby modifying quantum transport properties.
Among various microscopic mechanisms, electron--phonon (e--ph)
interaction is one of the primary sources of dephasing in realistic
systems. To incorporate such effects phenomenologically, B\"{u}ttiker
proposed an elegant approach based on virtual voltage probes connected
to each lattice site of the conducting region~\cite{Deph1,Deph2,Deph3,Datta1995,Datta2005}.
These probes do not carry net current but introduce phase randomization
of the electronic wavefunction, effectively simulating incoherent
scattering processes. To maintain the zero-current condition at the dephasing probes,
we assume a linear voltage drop along the chain, such that the electrochemical
potentials vary uniformly between the source and drain.~\cite{Patra2019}
Since this approach provides a convenient
and widely used framework to incorporate dephasing, we adopt it in the
present analysis to examine how phase-breaking processes influence the
transport behavior of edge, band-edge bulk, and in-band bulk states.

\begin{figure}
{\centering \resizebox*{8cm}{6cm}{\includegraphics{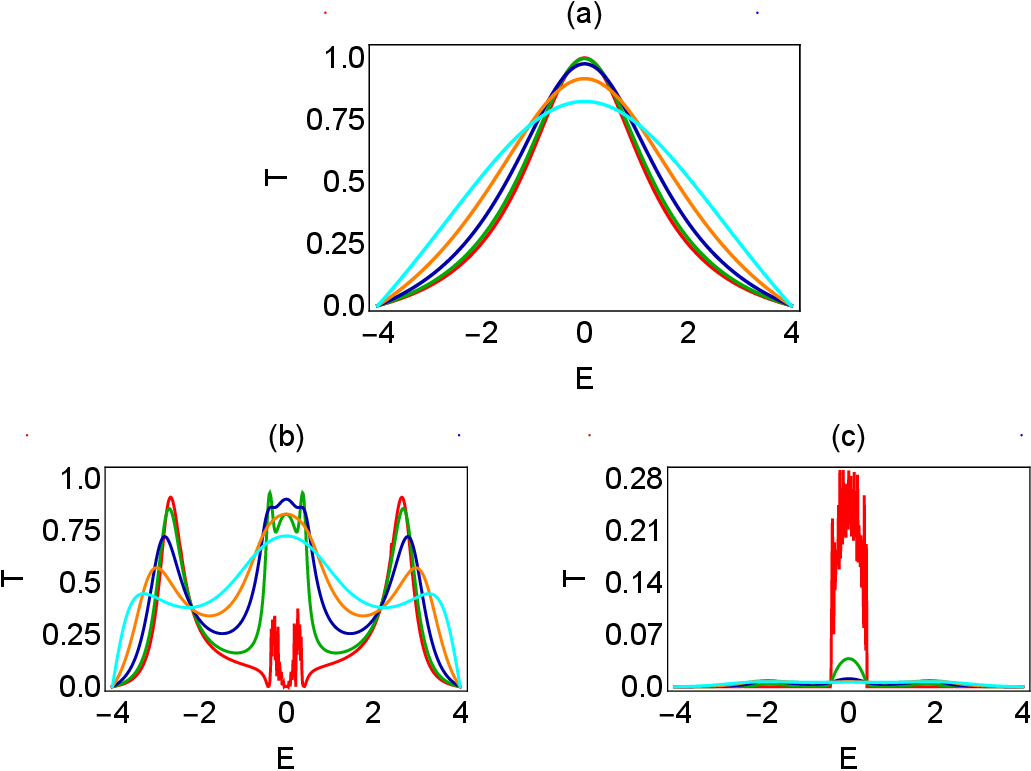}}\par}
\caption{(Color online). Transmission probability as a function of
energy $E$ for different dephasing strengths in an off-diagonal
AAH chain of length $N=60$, where B\"{u}ttiker probes are attached
to each lattice site. The red, green, blue, orange, and cyan curves
correspond to dephasing strengths of $0$, $0.3$, $0.6$, $0.9$, and
$1.2,$eV, respectively. (a) Edge-state transport with $N_S=N_D=1$
and $\phi=\pi/2$, showing weak sensitivity to small dephasing and
gradual peak suppression at larger dephasing strengths.
(b) Band-edge bulk-state transport with $N_S=4$, $N_D=5$, and
$\phi=\pi/4$, exhibiting moderate suppression and peak broadening.
(c) In-band bulk-state transport with $N_S=1$, $N_D=60$, and
$\phi=\pi/4$, showing strong suppression of transmission with increasing
dephasing.}
\label{f9}
\end{figure}

Conventionally, the coupling strength between the B\"{u}ttiker probes and
the chain is taken as a measure of the dephasing strength. In
Fig.~\ref{f9}, we present the transmission spectra for different values
of the dephasing strength. The red, green, blue, orange, and cyan curves
correspond to dephasing strengths of $0$, $0.3$, $0.6$, $0.9$, and
$1.2\,\mathrm{eV}$, respectively. In all cases, we consider a chain of
length $N=60$, with B\"{u}ttiker probes attached to every lattice site.

In Fig.~\ref{f9}(a), we examine the edge-state transmission spectra for
$N_S = N_D = 1$ and $\phi = \pi/2$. For small dephasing strengths, the
transmission remains almost unchanged, as the red and green curves are
nearly identical. With increasing dephasing strength, the peak heights
gradually decrease (reaching $T \sim 0.75$ at a dephasing strength of
$1.2\,\mathrm{eV}$), while the peaks broaden, indicating enhanced phase
randomization. This weak sensitivity to dephasing can be attributed to
the localized nature of edge states, which require only limited phase
coherence for transport.

In Fig.~\ref{f9}(b), we investigate the effect of dephasing on band-edge
bulk states for $N_S = 4$, $N_D = 5$, and $\phi = \pi/4$. We observe a
moderate suppression of the transmission (from $T \sim 0.9$ to
$T \sim 0.45$), accompanied by peak broadening, reflecting the partially
localized character of these states. Interestingly, the in-band bulk
states around $E = 0$, which exhibit relatively small transmission
($T \sim 0.28$ with a minimum near $E = 0$) in the absence of dephasing,
show a noticeable enhancement ($T \sim 0.8$) for weak dephasing before
undergoing gradual suppression at stronger dephasing. This behavior
suggests that weak dephasing partially suppresses destructive quantum
interference, thereby facilitating transport, whereas stronger dephasing
progressively destroys phase coherence and reduces the transmission.

Finally, in Fig.~\ref{f9}(c), we analyze the in-band bulk states around
$E = 0$ for $N_S = 1$, $N_D = 60$, and $\phi = \pi/4$. In this
configuration, we observe a strong suppression of the transmission with
increasing dephasing, where $T$ decreases significantly from
$T \sim 0.28$ to $T \sim 0.006$ as the dephasing strength increases from
$0$ to $1.2\,\mathrm{eV}$. This pronounced reduction highlights the strong
dependence of extended bulk states on phase coherence, as transport in
these states relies on coherent propagation across the entire system.

This behavior indicates that weak dephasing can partially suppress
destructive interference and thereby enhance transmission, whereas strong
dephasing progressively destroys phase coherence and suppresses coherent
transport.

\section{Experimental Realization}

The transport properties predicted in this work can be experimentally
realized in controllable platforms such as photonic waveguide arrays,
ultracold atoms in optical lattices, or semiconductor quantum dot and
nanowire structures, where modulated hopping amplitudes and finite
one-dimensional chains can be engineered with high
precision~\cite{Rechtsman2013,Verbin2013,Roati2008,Schreiber2015,
Hanson2007}. Site-selective contacts in such systems enable the
investigation of electrode-configuration–dependent transport, while
controlled coupling to external reservoirs or environmental noise can be
used to emulate dephasing effects~\cite{Buttiker1986,Datta1995,
Luschen2017}. Therefore, the key features reported here, including the
distinct transport signatures of edge, band-edge bulk, and in-band bulk
states, should be experimentally accessible.

\section{Conclusion}

We have studied transport in an off-diagonal Aubry--André--Harper chain,
focusing on the interplay between quasiperiodic modulation, system size,
electrode configuration, and dephasing. The modulation generates effective
internal boundaries, which play a central role in shaping both spectral
and transport properties.

We identify distinct transport signatures of edge, band-edge bulk, and
in-band bulk states. Edge-state transport is robust and exhibits a
system-size periodicity governed by the family structure $N = 4m + r$.
In contrast, in-band bulk states show strong finite-size interference
effects, while band-edge bulk states display intermediate behavior with
smoother transmission characteristics.

Using the participation ratio, we find a crossover in the zero-energy
state from extended to localized behavior with increasing modulation
strength, eventually leading to size-independent localization.

Dephasing effects further distinguish these states: edge states are
robust against weak dephasing, band-edge bulk states are moderately
suppressed, and in-band bulk states are strongly affected due to their
reliance on coherent transport.

Overall, the results demonstrate that internal boundaries, electrode
geometry, and environmental decoherence provide effective control over
transport in quasiperiodic systems, with clear and experimentally
accessible signatures in photonic, cold-atom, and semiconductor
platforms.

\section{Acknowledgement}

The author gratefully acknowledges Prof. Santanu K. Maiti for teaching
the fundamental concepts and physical insights that formed the basis for
this work.

\end{document}